\documentclass{emulateapj}
\usepackage{natbib}
\newcommand{\uv}{\mbox{$u$-$v$}}
\newcommand{\ex}[1]{\mbox{$\times 10^{#1}$}}

\newcommand{\muas}{\mbox{$\mu$as}}
\newcommand{\muasyr}{\mbox{$\mu$as yr$^{-1}$}}
\newcommand{\masyr}{\mbox{mas~yr$^{-1}$}}

\newcommand{\kms}{\mbox{km s$^{-1}$}}

\newcommand{\Jb}{\mbox{Jy bm$^{-1}$}}
\newcommand{\Ra}[4]{\mbox{${#1}^{\rm h} \; {#2}^{\rm m} \; {#3}\fs{#4} $}}
\newcommand{\dec}[4]{\mbox{${#1}\arcdeg \; {#2}\arcmin \; {#3}\farcs{#4} $}}

\shortauthors{Bietenholz \& Bartel}
\shorttitle{SN 2001em: No GRB Event}

\defcitealias{SN2001-1}{Paper I}

\begin{document}
      
\title{SN 2001\lowercase{em}: No Jet-Driven Gamma Ray Burst Event}

\author{M. F. Bietenholz\altaffilmark{1,2} and N. Bartel\altaffilmark{2}}

\altaffiltext{1}{Hartebeesthoek Radio Observatory, PO Box 443, Krugersdorp, 
1740, South Africa} 
\altaffiltext{2}{Department of Physics and Astronomy, York University, Toronto,
M3J~1P3, Ontario, Canada}
 
\slugcomment{{\em Accepted for publication in} Astrophysical Journal Letters}

\begin{abstract}
We report on our second-epoch VLBI and VLA observations of the Type
Ib/c supernova 2001em, five years after the explosion.  
It was suggested that SN~2001em might be a jet-driven
gamma ray burst (GRB), with the jet oriented near the plane of the
sky, which would entail relativistic expansion or motion.  Our VLBI
image shows that SN~2001em is still unresolved five years after the
explosion.  For a distance of 83~Mpc ($H_0$ = 70 \kms~Mpc$^{-1}$), the
nominal expansion velocity is $5800 \pm 10,000$~\kms, and the proper
motion is $33,000 \pm 34,000$~\kms.  Our values are inconsistent with
either relativistic expansion or motion, but are consistent with the
non-relativistic expansion speeds and small proper motions
seen in other supernovae.  In particular these values are consistent with
radio emission from SN~2001em being due to normal, non-relativistic
supernova ejecta interacting with the circumstellar medium. Our VLA
observations show a power-law decay in flux density since the time of
the peak in the 8.4~GHz radio lightcurve in $\sim$2003.
\end{abstract}

\keywords{supernovae: individual (SN2001em), radio continuum: general,
gamma rays: bursts}

\section{INTRODUCTION}

Supernova 2001em was discovered on 2001 September 15
\citep{PapenkovaL2001}, in the galaxy \objectname{UGC 11794}
(\objectname{NGC 7112}; $z = 0.01935$, $D = 83$~Mpc for $H_0$ = 70
\kms~Mpc$^{-1}$).  It was classified as a Type Ib/c on the basis of an
early optical spectrum \citep{FilippenkoC2001}.  Subsequently,
however, it went on to show strong signs of circumstellar interaction
\citep[see][and references therein]{ChugaiC2006}.  Among these signs
were radio and X-ray luminosities unprecedented for a Type Ib/c
supernova after $\sim$2~yr \citep{Stockdale+2004, Stockdale+2005,
PooleyL2004}.

It was suggested that the radio emission from \objectname{SN 2001em}
might be due to the interaction with the circumstellar medium (CSM) of
a relativistic jet, oriented near the plane of the sky \citep[][see
also Paczy\'nski 2001]{GranotR2004}\nocite{Paczynski2001}.  Such jets
are thought to power gamma-ray burst (GRB) events; in particular long
duration GRB events, which seem to be associated with Type Ib/c
supernova.

A prediction of the relativistic jet model for SN~2001em is that the
source should be resolvable by very long baseline interferometry
(VLBI).  Along with two other groups, we undertook such observations
\citep[] [referred to hereafter as \citetalias{SN2001-1}]{SN2001-1}\@.
The two other groups reported their results in \citet{Stockdale+2005}
and \citet{Paragi+2005}.  SN~2001em was at best marginally resolved in
all three experiments.  We obtained the strongest limit on the
expansion speed, namely a $3\sigma$ limit of 70,000~\kms\ on one-sided
expansion \citepalias{SN2001-1}, which does not support the
relativistic jet model.  An alternate model was proposed by
\citet{ChugaiC2006}, which involves the interaction of normal
(non-relativistic) Type~Ib/c ejecta with a massive and dense
circumstellar shell, produced by mass loss of the progenitor.

The relativistic jet model, however, is not entirely excluded by the
small size observed for the radio emission region:  it is possible
that only the ``working end'' of the relativistic jet is
radio-bright.  In this case, however, the radio-bright spot should
display a relativistic proper motion.  At 83~Mpc, a speed of $c$
corresponds to $0.8$~\masyr, which is resolvable by current VLBI
arrays in less than one year.

To better measure the expansion speed, and to determine the proper
motion of SN~2001 we undertook ``second epoch'' VLBI observations in
2006 May, 4.7~yr after the explosion, and 1.4~yr after our previous
observations.  We report the results of these observations here.

\section{OBSERVATIONS}

As in \citetalias{SN2001-1}, we observed SN~2001em using the ``High
Sensitivity Array,'' which consists of NRAO\footnote{The National
Radio Astronomy Observatory, NRAO, is operated under license by
Associated Universities, Inc., under cooperative agreement with
National Science Foundation, NSF.}'s Very Long Baseline Array (VLBA;
each of 25~m diameter), phased Very Large Array, (VLA; 130~m
equivalent diameter), and Robert C. Byrd telescopes ($\sim$105~m
diameter), in addition to the Arecibo Radio Telescope\footnote{The
Arecibo Observatory is part of the National Astronomy and Ionosphere
Center, which is operated by Cornell University under a cooperative
agreement with the NSF.} (305~m diameter) and the Effelsberg Radio
Telescope\footnote{The 100~m telescope at Effelsberg is operated by
the Max-Planck-Institut f\"{u}r Radioastronomie in Bonn, Germany.}
(100~m diameter).  The VLBI observations were carried out on 2006 May
27, with a total time of 12~hrs.  We observed at a frequency of
8.4~GHz with a bandwidth of 32~MHz and recorded both senses of
circular polarization with a total recording bit rate of
256~Mbits~s$^{-1}$. The data were correlated with the NRAO VLBA
processor at Socorro.  The analysis was carried out with NRAO's
Astronomical Image Processing System (AIPS)\@.  The initial flux
density calibration was done through measurements of the system
temperature at each telescope, and then improved through
self-calibration of the reference sources. Finally, in addition to
using the phased VLA, as an element in our VLBI array, we also used
the VLA, which was in the AB configuration, as a stand-alone
interferometer to determine the integrated flux density of SN~2001em
at 1.5, 8.4 and 22~GHz.

We phase-referenced our VLBI observations to the same two sources we
used in \citetalias{SN2001-1}, namely \objectname{JVAS J2145+1115} and
\objectname{JVAS J2139+1423}.  For most of the observing session, we
phase-referenced to our primary phase-reference source, J2145+1115,
which is 1.4\arcdeg\ away on the sky.  In addition, we had several
$\sim$20~min periods where we phase-referenced to our secondary source
J2139+1423, which is 2.1\arcdeg\ away on the sky, and is a ``defining
source'' of the International Celestial Reference Frame
\citep[ICRF;][]{Ma+1998}.  The use of the second reference source
provides both a check on the astrometry and on possible source motions
of J2145+1115, and allows us to determine coordinates for SN~2001em
tied to the ICRF\@.  For the observations, we used a cycle-time of
$\sim$240~s, with $\sim$170~s spent on SN~2001em and $\sim$70~s spent
on the calibrator, either J2145+1115 or J2139+1423.

\section{RESULTS}

\subsection{VLA Total Flux Densities and the Radio Lightcurve}

Using the VLA as a stand-alone interferometer, we measured the integrated
flux density of SN~2001em on 2006 May 27 to be 1.05~mJy at both 1.4
and 8.4~GHz, with uncertainties of 0.06 and 0.22~mJy, respectively.
At 22.5~GHz, we did not detect SN~2001em, and instead place a
$3\sigma$ upper limit on its flux density of 0.8~mJy.  We show the
radio lightcurve of SN~2001em in Fig.~\ref{flightcurve}.  The
lightcurve reaches a peak at time after the explosion, $t$, of
$\sim$2.8~yr, and shows an exponential decay thereafter, with the flux
density being proportional to $t^{-1.05 \pm 0.02}$.

\subsection{Astrometric Results: the Proper Motion}

The astrometric results were obtained by fitting models in the
\uv~plane to the phase-referenced visibility data from the 2006 May as
well as the 2004 November observations.  Arecibo was not used for
deriving the astrometric results, as systematic phase-shifts between
the calibrator sources and SN~2001em were observed on baselines
involving Arecibo.  For the 2004 epoch, we also did not use data from
the Kitt Peak VLBA station, as they were of poor quality.  For data
from the remainder of the array, we treated the sections of data
phase-referenced to J2145+1115 separately from those phase-referenced
to J2139+1423.  The model-fitting involves the least-squares fitting
of geometrical models to the VLBI visibility data in the Fourier
transform or \uv\ plane, and is described in more detail in, e.g.,
\citet{SN93J-1} and \citet{SN93J-2}.

We fit a model consisting of a circular Gaussian to the fully
calibrated visibility data.  Since the source is only marginally
resolved, the choice of model has little effect on our astrometric
results. For example, fitting either a point-source, an elongated
elliptical Gaussian, or a spherical shell model produces the same
center position to well within the uncertainties.  The statistical
uncertainties in the fit positions are on the order of 50~\muas, but
the true uncertainties are likely dominated by systematic effects,
such as those of unmodelled parts of the troposphere and ionosphere
and evolution of the reference sources.  We will therefore use an
estimated $1\sigma$ uncertainty of 100 and 280~\muas\ in each
coordinate when phase-referencing to J2145+1115 and J2139+1423,
respectively. The uncertainties when phase-referencing to J2139+1423
are higher both because it is farther away on the sky and because we
spent only a fraction of our observing time using it as the reference
source.  For further discussion of the limitations on the astrometric
accuracy obtainable with VLBI, see, e.g., \citet{SN93J-1, M81-2004,
PradelCL2006}.

If we use the larger segment of our data in which we phase-referenced
to J2145+1115 we find a proper motion of SN~2001em relative to that
source of $-66 \pm 93$ and $60 \pm 93$~\muasyr\ in RA and decl.,
respectively, or $89 \pm 93$~\muasyr\ in total (Fig.~\ref{fimg}
shows the locations of the fit positions for both epochs on the
2006 image).
If, instead, we use the smaller portion of our data, in which we
phase-reference to our second reference source, J2139+1423, we
obtain a consistent, but less accurate, relative proper motion of
$192 \pm 280$ and $20 \pm 280$~\muasyr\ in RA and decl., respectively.

To what extent is our proper motion of SN~2001em dependent on the
stationarity of the reference sources?  Since the relative proper
motion between SN~2001em and J2145+1145 was low, it is likely that the
proper motion of J2145+1145 is also low, since it would be unlikely
that both SN~2001em and J2145+1145 have large proper motions which
fortuitously cancel in our differential measurement.  It is more
likely that the proper motion of SN~2001em is in fact consistent with
zero at the level of our uncertainties.  Our second reference source,
J2132+1423, is an ICRF source whose proper motion has been
measured to be $<40$~\muasyr\ \citep{Feissel-Vernier2003}.
Our measured proper motion of SN~2001em relative to it was also
consistent with zero to within $1\sigma$.  We therefore consider it
reasonable to assume that our primary reference source, J2145+1145, is
stationary, and that the proper motion of SN~2001em is
$$\mu = 89 \pm 93 \; \muasyr, \; ({\rm p.a.}=  -\!138\arcdeg).$$
At 83~Mpc, this proper motion corresponds to a non-relativistic speed of 
$$v = 33,000 \pm 34,000 \; \kms, \; ({\rm p.a.}= -\!138\arcdeg).$$
We note that even our less accurate
value of the proper motion of SN~2001em with respect to J2139+1423 is
inconsistent with a projected speed of $c$ at the $2.2\sigma$ level.

\subsection{Images and the Size of the Radio-Emitting Region}

To obtain the best imaging results, we phase self-calibrated the data
for SN~2001em, now including the data for Arecibo and combining the
sections phase-referenced to J2145+1115 and J2139+1423.  This
self-calibration serves to correct the mentioned systematic
phase-shift seen at Arecibo, and also any phase shifts due to possible
inconsistencies in the positions of the two reference sources.

In Figure \ref{fimg} we show the VLBI image of SN~2001em on 2006
May 27.  No structure is apparent.  The source is, as it was in our
2004 observations, at best marginally resolved at our resolution of
$1.7 \times 0.8$~mas at p.a.\ $-1$\arcdeg\ (FWHM of a Gaussian; p.a.\
is east of north). In particular, no structure larger than $\sim
1.7$~mas or $\sim 2\ex{18}$~cm is visible.

To get a more accurate determination of the size, we turned again to
model fitting in the \uv~plane, which has the advantage of giving
results independent of the convolving beam.  Since the source is at
best marginally resolved, we must choose the model {\em a priori}. We
use the same models as in \citetalias{SN2001-1}, namely an elliptical
Gaussian, which we choose to represent a possibly elongated source
such as a jet, and a spherical shell model appropriate for a
supernova.  The shell model consists of the projection of a
spherically symmetrical, optically thin shell of emission, with the
shell width being 20\% of the outer radius \citep[as was found
appropriate for SN~1993J; see e.g.][]{SN93J-3}.

We discussed the uncertainties in the fitting process in detail in
\citetalias{SN2001-1}, \citep[see also][]{SN93J-1, SN93J-2}.  To
summarize, the uncertainties are likely dominated by residual
inaccuracies in the complex telescope gains.  We again use
the square root of the visibility weights in our fits, which results
in a more robust fit less dominated by a few very sensitive baselines.
We estimated the uncertainties by allowing the antenna gains to vary
and by comparing the fitted values with and without Arecibo, our most
sensitive antenna.

Using an elliptical Gaussian model appropriate for a jet, we obtained
a FWHM major axis size of $0.16 \pm 0.18$~mas, corresponding to a
$3\sigma$ upper limit of 0.7~mas.  At 83~Mpc, this limit corresponds
to a size of $8 \ex{17}$~cm, assuming undecelerated, one-sided,
motion since the explosion, to a speed of 54,000~\kms.
For the spherical shell model appropriate for a supernova, we find an
outer angular radius of $0.07 \pm 0.12$~mas, with a
corresponding $3\sigma$ upper limit of 0.42~mas\footnote{We note that
other models with a geometry different from a spherical shell will
give different outer radii, but for likely geometries, the differences
to our selected shell model are $<25$\%. For example, a uniform sphere
model gives a radius $\sim 16$\% larger, whereas a very thin shell
model gives a radius $\sim 10$\% smaller \citep[see,
e.g.,][]{SN93J-2}.}.  At 83~Mpc, this corresponds to a size of $(9 \pm
15) \ex{16}$~cm and an average expansion velocity (since the
explosion) of $5800 \pm 10,000$~\kms.

\section{DISCUSSION}

We have made a second-epoch VLBI image of SN~2001em, one of the
first Type Ib/c supernova to be observed with VLBI.  SN~2001em had an
unusually delayed onset of radio emission, and it had been suggested
that the radio emission might be due to a relativistic jet not aligned
with the line of sight, perhaps linking SN~2001em to a GRB
\citep{GranotR2004}.  However, an alternate model, where the radio
emission is due to the normal Type Ib/c ejecta overtaking a dense
shell left by an episode of strong mass-loss of the progenitor, can
also reproduce the late-onset radio emission \citep{ChugaiC2006}.

\begin{figure}
\epsscale{0.75}
\plotone{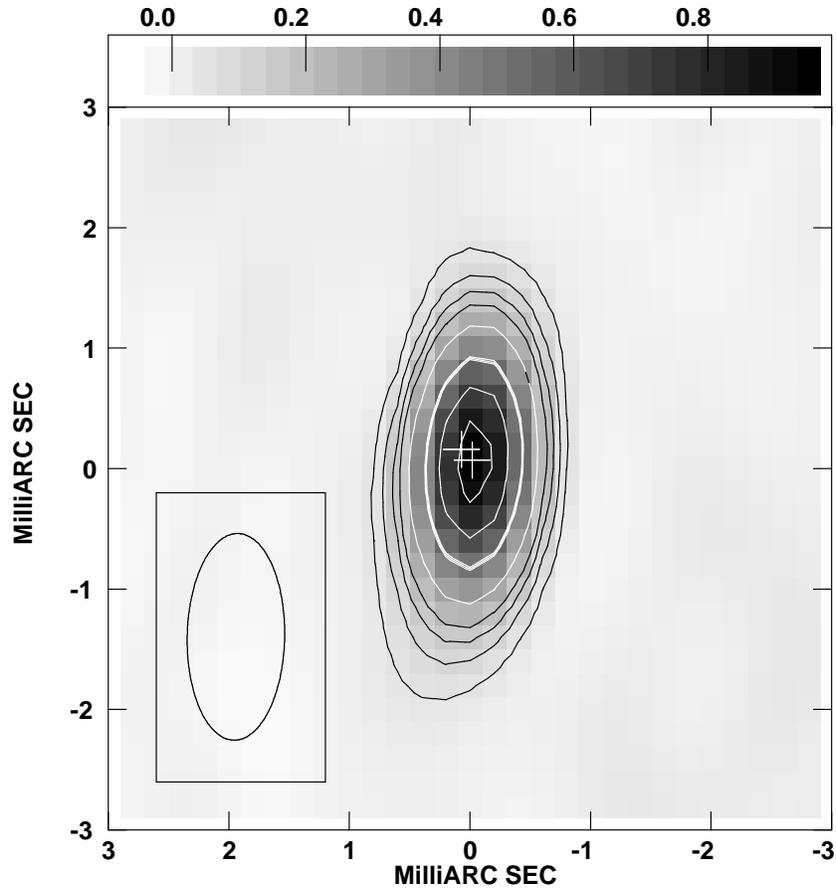}
\caption{
A VLBI image of SN~2001em on 2006 May 27.  The peak brightness was
950~$\mu$\Jb\ and the rms background noise was 15~$\mu$\Jb.  The
contours are drawn at $-5$, 5, 10, 20, 30, {\bf 50}, 70 and 90\% of
the peak brightness, with the lowest contour being at $3\sigma$, and
the 50\% contour being emphasized. The FWHM size of the Gaussian
restoring beam, which can be compared to the 50\% contour of the
source, was $1.7 \times 0.8$~mas at p.a.\ $-1$\arcdeg, and is
indicated at lower left.  The gray scale is labelled in $\mu$\Jb.  The
two crosses show the fit position for the two epochs, with the right
cross representing the current epoch (2006.4) and the left one the
2004.9 epoch.  The origin is at R.A. = \Ra{21}{42}{23}{60938}, decl.\
= \dec{12}{29}{50}{3001} (J2000).  North is up and east to the left.}
\label{fimg}
\end{figure}

Our present VLBI image (Fig.~\ref{fimg}) shows that SN~2001em is still
only marginally resolved 4.7~yr after the explosion.  The radio
emission region is not expanding relativistically.  We also measured
the proper motion of the radio emission region between our 2004 and
2006 epochs, and found no significant proper motion.  The 3$\sigma$
upper limit on the source's velocity in the plane of the sky was
135,000~\kms\ or $0.45\,c$.  The radio-bright region, therefore, does
not exhibit any relativistic projected motion, but rather is
consistent with being stationary.  Different viewing angles can
produce a wide range of projected speeds even if the true speed is
near $c$.  We can, however, constrain the range of viewing angles for
a standard jet model where the apparent jet velocity $\beta_{\rm
app}$, the true jet velocity $\beta$ (both in units of $c$), and the angle
$\theta$ between the jet and our line of sight are related through
$\beta_{\rm app} = \beta \sin{\theta}/(1-\beta \cos{\theta})$
\citep[e.g.,][]{PearsonZ1987}.
For example, if $\beta = 0.95$, then only a jet oriented at either
$\theta >130\arcdeg$ or $\theta <1.4\arcdeg$ will $\beta_{\rm app} <
0.45$.  Thus, if we assume we are viewing the Doppler-boosted
approaching jet, the only way to reconcile a true velocity $>0.95\,c$
with our proper motion results is to assume that the jet is
fortuitously aligned to within 1.4\arcdeg\ of the line of sight.

The fact that the radio-emitting region is neither expanding nor
moving relativistically therefore argues against the presence of a
relativistic jet in SN~2001em.  A further argument against such a jet
is the detection from SN~2001em of strong H$\alpha$ emission with a
full width at half maximum of 1800~\kms\ \citep{SoderbergGK2004}. Such
emission is usually observed in supernovae undergoing strong
circumstellar interaction.

Both the radio morphology and proper motion of SN~2001em, on the other
hand, are consistent with the hypothesis of non-relativistic supernova
ejecta interacting with the CSM, as was proposed by
\citet{ChugaiC2006}.  In this model, the late onset radio and X-ray
emission is caused by the interaction of the supernova ejecta with a
dense shell in the CSM caused by an episode of rapid mass loss of the
progenitor.
This model can naturally account for the H$\alpha$ emission, which is
probably circumstellar gas accelerated by the forward shock.

\begin{figure}
\epsscale{0.6}
\plotone{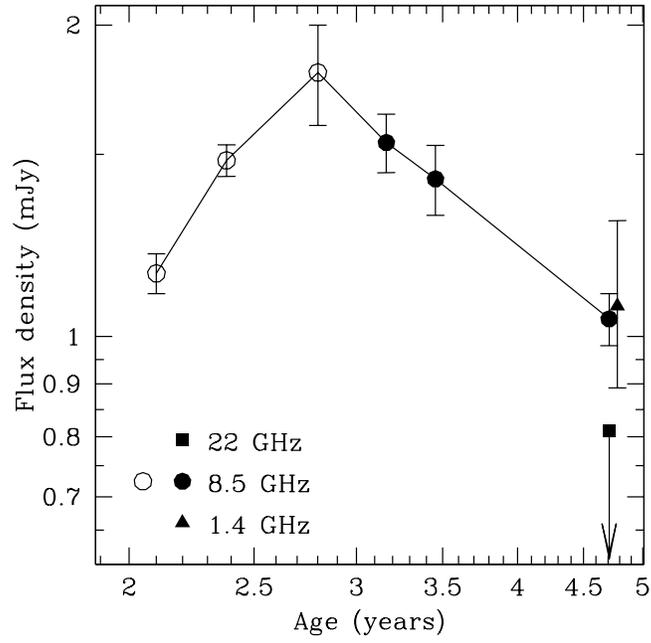}
\caption{The radio lightcurve of SN~2001em at 8.4~GHz, with our
additional flux density measurement at 1.4~GHz (shifted slightly for
clarity) and our $3\sigma$ upper limit on the radio flux density at
22~GHz, both for 2006 May.  The data points are from VLA measurements
except that at $t = 2.8$~yr, which is from the VLBI observations.  The
error-bars are standard errors consisting of the statistical uncertainty
and that in the flux density calibration added in quadrature.  The
filled points are our own measurements and the open circles are
published values from \citet{Stockdale+2005, Stockdale+2004}.}
\label{flightcurve}
\end{figure}

When we fit a spherical shell model to the radio emission, we found a
nominal size in 2006.4 of $(9 \pm 15)\ex{16}$~cm and an average
expansion speed since the explosion of $5800 \pm 10,000$~\kms.
Although our estimates are rather uncertain, both are compatible with
the typical values usually seen in supernovae
\citep[e.g.,][]{VLBA10th}, and in particular with the values used in
the Chugai \& Chevalier model for SN~2001em.   We note
that VLBI observations were also made of Type Ib/c SN~2003L,
however only a rather large limit of $5.4 c$ was obtained for
directly measured expansion velocity \citep{Soderberg+2005}.
In addition, the radio lightcurve (Fig.~\ref{flightcurve}) appears to
exhibit a power-law decay with flux density $\propto t^{-1.05}$.  This
is shallower than that predicted for the jet model
\citep{Granot+2002}, but such shallow decays are seen in some Type~II
radio supernova \citep[e.g.,][]{SN79C}.  As an aside, we note that our
expansion, proper-motion, and flux density measurements are compatible
also with a young pulsar nebula, as is possibly seen in SN~1986J
\citep{SN86J-CosparII, SN86J-Sci}, although a pulsar nebula would not
be expected to produce the observed broad H$\alpha$ emission.

In summary, we obtained second epoch VLBI observations of the
Type~Ib/c supernova 2001em, which had unusually late turn-on radio and
X-ray emission.  We find no evidence of relativistic expansion or
motion of the radio emitting region.  Rather, we find that the proper
motion is consistent with SN~2001em being stationary and that the
expansion speed is consistent with those of normal supernovae.  These
results favour a model where normal Type~Ib/c supernova ejecta are
interacting with a dense circumstellar shell.

\acknowledgements 
\noindent\mbox{Research at York University was partly supported by NSERC}.

\bibliographystyle{apj}
\bibliography{mybib1}

\clearpage

\end{document}